%% file: ms.tex
\documentclass[12pt,preprint]{aastex}

\begin{document}


\title{Atomic jets from class 0 sources detected by Spitzer: the case of L1448-C}

\author{O. Dionatos, B. Nisini, R. Garcia Lopez, T. Giannini}
\affil{INAF - Osservatorio Astronomico di Roma Via di Frascati 33,
00040, Monteporzio Catone, Italy}
\email{dionatos@oa-roma.inaf.it, nisini@oa-roma.inaf.it,garcia@oa-roma.inaf.it,giannini@oa-roma.inaf.it}

\author{C. J. Davis}
\affil{Joint Asronomy Centre, 660 North A'ohoku Place, University Park Hilo, Hawaii 96720, U.S.A.}
\email{c.davis@jach.hawaii.edu}

\author{M. D. Smith}
\affil{ School of Physical Sciences, Ingram Building, The University of Kent,
Canterbury CT2 7NH, U.K.}
\email{m.d.smith@kent.ac.uk}

\author{T. P. Ray}
\affil{Dublin Institute for Advanced Studies, 31 Fitzwilliam Place, Dublin 2,
Ireland }
\email{ tr@cp.dias.ie}

\author{M. De Luca}
\affil{Indiana University Cyclotron Facility,
2401 N. Milo B. Sampson Lane
47408 Bloomington, IN
U.S.A}
\email{delucama@indiana.edu}

\begin{abstract}
We present Spitzer-IRS spectra obtained along the molecular jet from the Class 0 source L1448-C (or L1448-mm). 
Atomic lines from the fundamental transitions of [FeII], [SiII] and [SI] have been detected showing, for the first time, the presence of an embedded atomic jet at low excitation. Pure rotational H$_2$ lines are also detected, and a decrease of the   atomic/molecular emission ratio is observed within 1$\arcmin$ from the driving source. 
Additional ground based spectra (UKIRT/UIST) were obtained to further constrain the H$_2$ excitation along the jet axis and,  combined with the 0--0 lines, have been compared with bow-shock models.
From the different line ratios, we find that the atomic gas is characterized by an electron density n$_e \sim$ 200-1000 cm$^{-3}$, a temperature $T_e <$ 2500 K and an ionization fraction $\lesssim$ 10$^{-2}$; 
the excitation conditions of the atomic jet are thus very different from those found in more evolved Class I and Class II jets.
We also infer that only a fraction (0.05-0.2) of Fe and Si is in gaseous form, indicating that dust still plays a major role in the depletion of refractory elements. A comparison with the SiO abundance recently derived in the jet from an analysis of several SiO sub-mm transitions, shows that the Si/SiO abundance ratio is $\sim$100, and thus that most of the silicon released from grains by sputtering and grain-grain collisions remains in atomic form. Finally, estimates of the atomic and molecular mass flux rates have been derived: values of the order of $\sim$10$^{-6}$ and $\sim$10$^{-7}$ M$_{\sun}$\,yr$^{-1}$ are inferred from the [SI]25$\mu$m and H$_2$ line luminosities, respectively. A comparison with the momentum flux   
of the CO molecular outflow suggests that the detected atomic jet has the power to drive the large scale outflow.

\end{abstract}


\section{Introduction}

The process of mass accretion, leading to the formation of solar type
stars, is always associated with mass ejection in the form of
collimated jets, which extend from a few AU up to parsecs away from
the exciting source. According to the models \citep{Koenigl, casse}, accretion and ejection are 
intimately related through the presence of a magnetized accretion
disk: the jets carry away the excess angular momentum, so that part of
the disk material can move towards the star. This paradigm of star
formation is now being observationally tested in Class I and Class II objects 
through detailed optical and near-IR observations \citep{ray}.

However, the characteristics of jets from these evolved  YSOs are unlikely to be appropriate 
 for those from protostars in earlier evolutionary phases, which
are expected to propagate in a denser medium and be associated with
more energetic mass ejection.  In such un-evolved objects, so-called Class 0 sources, the initial part of the jet is often
detectable only at mm wavelengths in the form of a collimated,
high-velocity molecular outflow \citep{gueth, lee, claudio}.  Near-IR H$_2$ emission is
always obscured by high extinction near the central engine and jet
base, and instead traces hot gas excited further downstream in bow
shocks near or at the jet apex. While it is usually assumed that the molecular jet represents the cold external layer of an embedded atomic jet, in principle the jet could be composed entirely of low excitation molecular gas and this needs to be tested observationally. 

Millimeter interferometric observations of high velocity molecular
jets from Class 0 sources present characteristics very
similar to the hot jets seen in T-Tauri or Class I YSOs, such as a
very narrow ($<$ 2$\arcsec$) width and a knotty structure resembling that
of HH objects \citep{gueth, sylvie}. Sub-mm observations in CO and SiO show that such jets
are dense, with peak values of n(H$_2$) $\sim$10$^6$ cm$^{-3}$ 
\citep[e.g][]{Bruni6}. Moreover, ISO-LWS observations suggest that
they are also warm, with temperatures between 300 and 1500 K inferred
from the copious high-J CO and H$_2$O emission \citep[]{teresa1}. 
Such a warm gas component may
represent the bulk of the mass flux ejected by the protostar, and
thus, energetically, may be the jet's most important component. The
very low LWS spatial resolution, however, has not permitted us to draw
any conclusions about the detailed structure of this warm gas, giving
only physical parameters averaged over the entire outflow.

Spitzer observations now allow us to investigate the properties of the warm gas component of
the molecular jet  through mid-IR atomic and molecular features probing low excitation gas
at T$\sim$100-2000 K and $n \sim$10$^4$-10$^6$ cm$^{-3}$.
The moderate IRS spatial resolution (4-20$\arcsec$) allows one to separate, in nearby sources,  the
inner jet region from the region where the jet is strongly interacting with the ambient medium through bow-shocks.

In this study, we present results of an analysis performed on Spitzer-IRS spectra of the jet from the Class 0 source L1448-C (or L1448-mm). This is a low luminosity \citep[$L$=7.5 L$_\odot$,][]{Tobin} protostellar source located in the Perseus Molecular Cloud \citep[D$\sim$ 250 pc, e.g.][]{enoch}. 
A powerful and highly collimated flow is driven by this source, as testified by interferometric CO and SiO maps \citep{guill,bachiller2}. This highly collimated molecular jet is associated with a less collimated and energetic CO outflow, probably representing ambient swept-up material. Near-IR observations performed on the L1448-C outflow show that H$_2$ hot molecular gas is detected only at the bow shocks at a distance of $\sim$1$\arcmin$ from the central source; the underlying jet that is responsible for the shocks remains undetected at optical and near-IR wavelengths and has been traced at mm-wavelengths as collimated SiO emission \citep{guill}.
 Far-IR observations performed with ISO have shown the existence of  a warm gas component (T$\sim$1000 K, n$\sim$10$^{5}$ cm$^{-3}$) associated with the molecular jet, showing up in [OI]63$\mu$m, H$_2$, CO and H$_2$O pure rotational emission \citep{Bruni1, Bruni2}. 
Recent Spitzer IRAC and MIPS images have shown that the source is actually binary, with a separation of $\sim$7$\arcsec$ \citep{Jorg, Tobin}. The north/south binary components are called L1448-CN and CS \citep{Jorg} or L1448-mm A and B \citep{Tobin} respectively. We will use the \citet{Jorg} nomenclature
in this paper. Of the two sources, the CN component is  associated with the strong millimeter source recognized by interferometric observations as the outflow driving source  \citep{bachiller2}.

The paper is organized as follows: \S 2 describes the Spitzer observations and data reduction as well as additional near-IR observations obtained at the UKIRT telescope.  \S 3 and 4 contain our analysis of the observed  H$_2$ and fine structure lines in order to derive the main physical parameters of the jet. Finally, conclusions are presented in \S 5 .

\section{Observations and data reduction \label{analysis}}

\subsection{ SPITZER - IRS \label{SPITZER}}

\begin{figure}
\epsscale{0.8}
\plotone{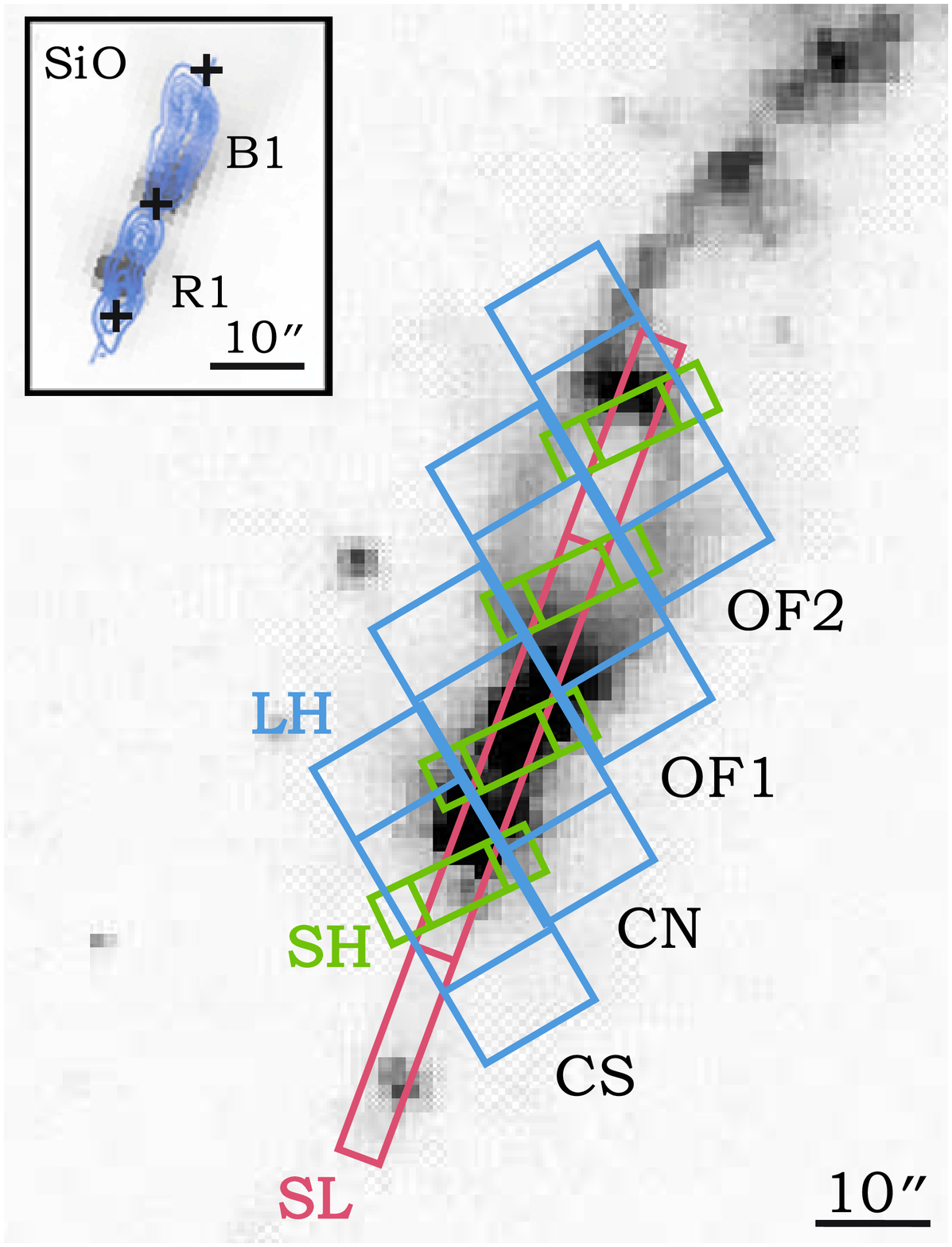}
\caption{Positioning of IRS slits for the SL, SH, \& LH modules, overlayed on 4.5 $\mu$m IRAC image (from the c2d survey). For the SL and LH modules, both nodding positions are displayed, whereas for the SH module only the outline of the observed area is presented for clarity.
The intersections of the slits define four areas of extraction along the jet axis. \textit{(Upper left)} SiO \citep{guill} 
contours overlayed on the 4.5 $\mu$m image show that the driving source is almost coincident with the northern emission area; crosses indicate the centers of the CS, CN and OF1 positions from bottom to top. 
\label{fig1}}
\end{figure}

The driving source and outflow of the Class 0 object L1448-C were observed with the SPITZER Infrared Spectrograph (Houck et al. 2004) during March and September 2006. 
Four positions have been targeted with the Short-High (SH), and Long-High (LH) IRS modules
($R \sim$ 600, 10-37$\mu$m) in order to cover the central driving source (L1448-CN), the southern CS source (the position of which spatially coincides with a CO red-shifted clump of Extremely High Velocity (EHV) jet emission named R1 \citep{bachiller1}, and two adjacent positions along the  blue-shifted part of the jet, that we call
OF1 and OF2, the first comprising the B1 mm clump in \citet{bachiller1} and  the second the H$_2$ bow-shock, as detected in the near-IR image of the flow by \citet{davis}, respectively  (see Figure \ref{fig1}). The plate scale of the detector is 2\farcs3 pixel$^{-1}$ and 
4\farcs5 pixel$^{-1}$ for the SH and LH modules, respectively. 
Observations with the Short-Low (SL) module ($R \sim$ 60-120, pixel scale 1\farcs8 pixel$^{-1}$) were also acquired to extend the spectral coverage down to 5$\mu$m. The observing period was chosen in order to have  the SL slit aligned along the L1448 jet. 
The observations were performed in staring mode with a total integration time of 2.5 hrs per position. 

The data were reduced and calibrated with the S13 pipeline. Basic calibrated data (bcd images) were median-combined 
and cleaned for rogue pixels with the IRSCLEAN$\_$MASK routine, and additional bad pixels in the low resolution module were removed by visual inspection. Spectral extraction was performed with the Spitzer IRS Custom Extraction tool (SPICE). For the high resolution modules the full slit was extracted, whereas for the SL module the full slit length was divided into four equal regions ($\sim$ 14$\arcsec$) that were consecutively extracted in order to best match the high resolution module pointings. 
High resolution spectra were defringed using the IRSFRINGE package while the zodiacal light contribution was estimated using the Spitzer Planning Observations Tool (SPOT) and subtracted. Inter-order flux offset and curvature effects were minimized by optimally selecting the best approximation between the point and extended source calibration options (SPICE), according to the morphology of each observed region. The resulting line flux difference between the two extraction methods is $\sim$ 20$\%$ which can be considered as an upper limit to the error for the observed line fluxes.

\begin{figure}
\epsscale{0.8}
\plotone{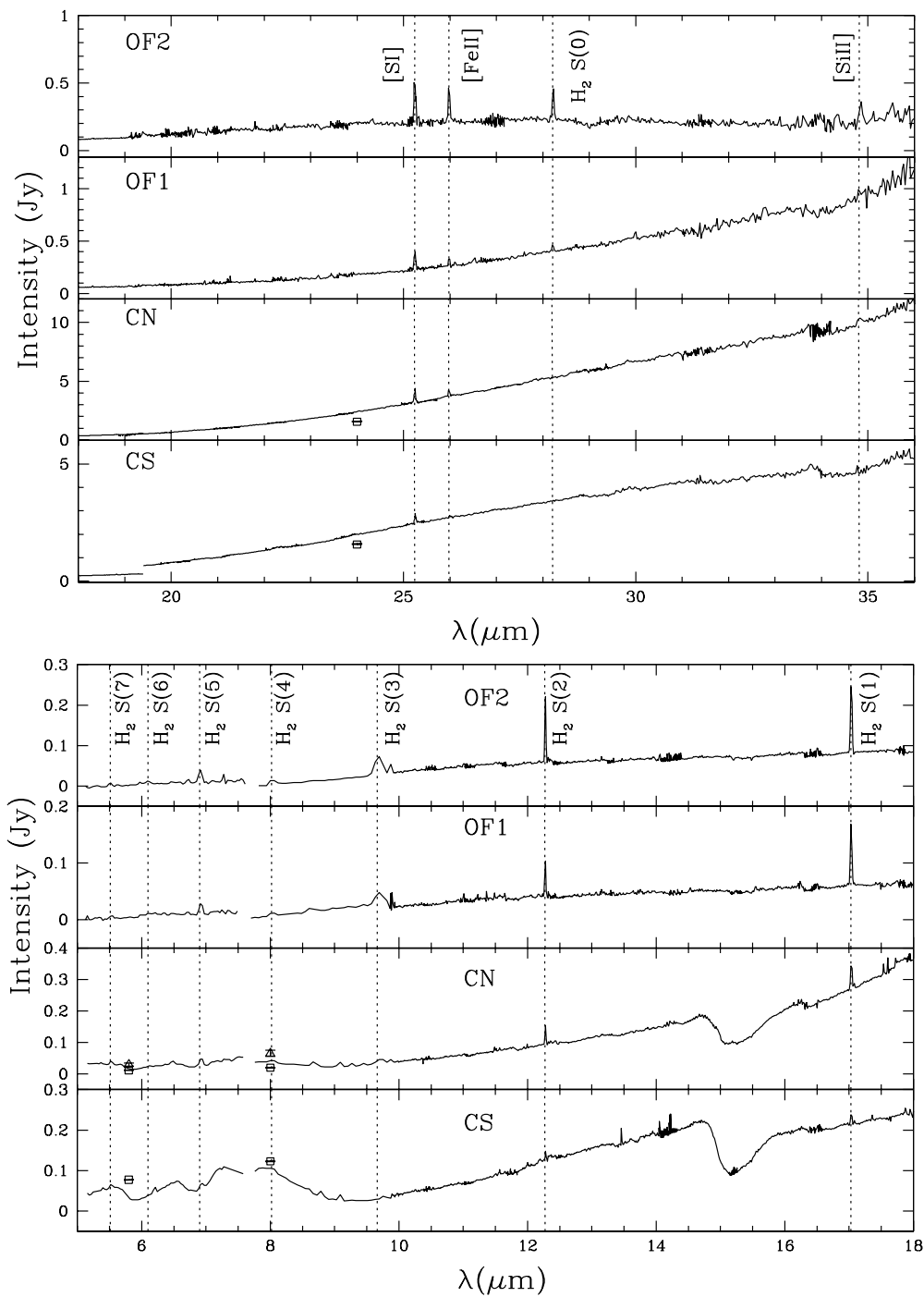}
\caption{Extracted spectra \textit{(lower panel: 5 - 18 $\mu$m, upper panel: 18 - 38 $\mu$m)} for each of the four positions defined in Figure \ref{fig1}, after defringing and removing zodiacal light. The observed lines are labeled on the top of each set. Open squares and triangles are IRAC bands 3,4 and MIPS band 1 continuum flux measuremens from \citet{Jorg} and \citet{Tobin}.
\label{fig2}}
\end{figure}

The calibrated spectra combined for all modules at each position are presented in Figure \ref{fig2}; in the same figure open squares and triangles represent IRAC bands 3,4 and MIPS band 1 continuum flux measurements from the recent studies of the region by \citet{Jorg} and \citet{Tobin} respectively. Their good agreement with the extracted spectra baseline confirms the appropriateness of the adopted extraction and calibration techniques. In particular, the procedure for zodiacal light subtraction is sufficiently precise to cancel the baseline jumps between different modules, the only exception being the SH and LH modules at the CS position, around 20$\mu$m. 

\begin{figure}
\epsscale{1.0}
\plotone{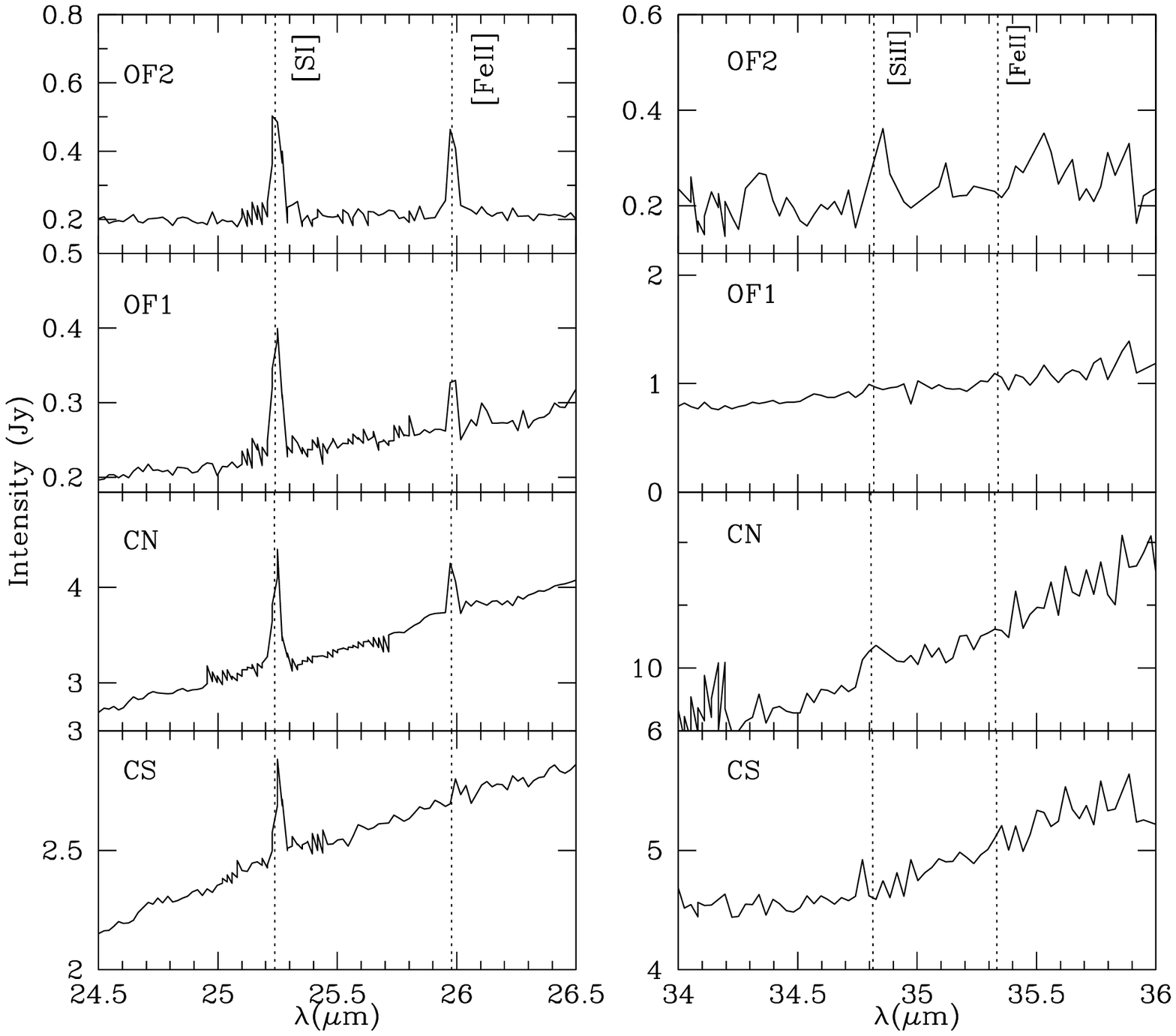}
\caption{Magnification of the spectra presented in Fig.\ref{fig2} displaying the ionic lines. 
\label{fig31}}
\end{figure}

The extracted spectra display a number of emission lines that arise both from molecular and forbidden atomic or ionic transitions. 
Line identification has been performed considering the features lying more than 3$\sigma$ above the local rms and whose emission peak, fitted by a gaussian, is within half a resolution element from the theoretical vacuum wavelength.
The observed molecular emission lines are due to H$_2$, where the full series of pure rotational transitions (S(0) - S(7)) in the given wavelength range of the instrument are detected.  
Atomic and ionic emission has been detected in the form of the fundamental fine-structure lines from  [FeII] (25.98$\mu$m), [SI](25.25$\mu$m) and [SiII] 34.81$\mu$m (see Fig.\ref{fig31}). None of the other numerous higher excitation  [FeII] lines falling in the investigated wavelength range has been detected at the 3$\sigma$ level: this fact, combined with the non-detection of the [NeII] line at 12.8 $\mu$m, already suggests that the investigated region presents very low excitation conditions.

In addition to emission lines, wide absorption bands are also observed in the spectra extracted at the two sources, 
the most prominent being those due to silicates at 9.7$\mu$m, water ice at 6.02$\mu$m and CO$_2$ ice at 15.2$\mu$m.

Fluxes of the individual lines were computed by gaussian fitting after subtracting a sloped baseline within the Spectroscopic Modeling, Analysis and Reduction Tool (SMART) \citep{higdon}.
 In Table \ref{tab1} we report the parameters of the detected lines, along with their identified transitions and the associated rms errors given by the  1\textit{$\sigma$} uncertainty derived from the  fluctuations of the local baseline. The measured fluxes have a larger uncertainty of up to 20\% given by the flux calibration uncertainty.
The 0--0 S(2) line has been observed by both the SL and SH modules. The line fluxes measured in the two modules agree inside the errors with the exception of the position OF2, where the SH module gives a flux higher by a factor of two. This difference could be due to a different bow-shock area sampled by the two modules. For
consistency, we use in all the positions the average of the two measurements.

In the same table we also give the upper limits for a set of non-detected lines that fall within the observed range and that will be used in the analysis that follows. 
In a previous study of  the same region with ISO/SWS, \citet{Bruni1} detected the S(3) to S(5) H$_2$ transitions in a region of 14$\arcsec \times 27\arcsec$ centered on the L1448-C source. The derived fluxes are a factor between 3 and 8 larger than the ones presented here. The FOV of the SWS observations is around 10 times larger than 
the region extracted in our spectra: therefore the observed flux difference can be attributed to the different adopted apertures.

\subsection{ UKIRT - UIST \label{UKIRT}}

In order to further investigate the excitation conditions along the jet axis, additional spectra, covering the
wavelength range from 1.4 to 2.5$\mu$m,  were obtained on 4th October 2006 at UKIRT (UK Infrared Telescope) using  the image-spectrometer UIST in spectroscopic mode. A 
4-pixel-wide slit was used with a pixel scale of 0.12$\arcsec$, corresponding to a spectral resolution 
$\sim$ 500-700. The slit was positioned roughly parallel with the jet axis and thus aligned with the SL IRS slit, at a position angle (PA) of -17$^o$.
Object-sky-sky-object sequences were made with a total exposure time of 300s for 
each one. Each spectral image was bias subtracted and flat-fielded using the 
ORAC data reduction software. Further reduction was performed using IRAF standard tasks. 
An Argon lamp was used in order to wavelength calibrate the spectra.
 B-type stars  were observed with the same configuration in order to remove 
the telluric features and flux calibrate the spectra.
Once the reduction was completed, the APALL  IRAF task was used to extract the
spectra corresponding in length to the four regions investigated with IRS.
Table \ref{tab2} lists the lines detected in the CN, OF1 and OF2  regions, with the corresponding measured flux. No lines have been detected in the CS position.
Only H$_2$ 1-0 and 2-1 ro-vibrational lines were detected, while any atomic or 
ionic emission (e.g. the strong [FeII] lines at 1.25$\mu$m and 1.64$\mu$m) is missing, as already evidenced by \citet{alessio}. 
It can be noted  that in the OF2 position, which corresponds to the bow-shock position, a 
large number of transitions  from higher excitation 1-0 lines (up to S(9)) are detected.
This may be indicative of  higher excitation conditions associated with the bow-shock or lower reddening
pertaining to this region since it is located far from the mm source core. 

\section{Analysis of the excitation conditions along the jet \label{discussion}}

\subsection{Extinction and spatial variation of line luminosity}
Optical extinction ($A_V$) values along the line of sight of the CS and CN sources have been measured from the 9.7$\mu$m silicate absorption feature, adopting the relationship between the optical depth of this feature and the visual extinction given by \citet{Mathis1}, i.e. $A_V/\tau_{9.7}$=19.3 mag. The derived $A_V$ values are 32 and 11 mag for the CS and CN source, respectively. In the OF1 and OF2 positions no silicate absorption is detected so no direct measurement of  $A_V$ from our spectra is possible. We have therefore assumed an extinction of 5 mag, as derived by \citet{Bruni2} towards the direction of the B1 clump (coincident with the OF1 position).
 Such a low value of extinction is in agreement with the absence of water and CO$_2$ ice absorption features in the spectra of these positions, as these features should become detectable in the Spitzer spectra only once the visual extinction reaches values above 4.3$\pm$1.0 mag \citep{Whittet}. All the individual detected lines have been dereddened adopting these $A_V$ values and the different extinction laws appropriate for the considered wavelength ranges: in the range from 1 to 13 $\mu$m the extinction law of \citet{R&L} was adopted while from 13 to 23  $\mu$m the extinction law of \citet{Mathis1} was used, which was extrapolated to 28 $\mu$m. 

An aposteriori check of the correctness of the adopted extinction values was done using the Boltzmann diagrams constructed from the H$_2$ near and mid-IR lines (see Sect. 3.2), by examining  the fit to a straight line of transitions appropriate for the 9.7$\mu$m silicate absorption (like the 0--0 S(3) line at 9.67$\mu$m) and at the interface between the near- and mid-IR lines in the same plot.

\begin{figure}
\epsscale{1.0}
\plotone{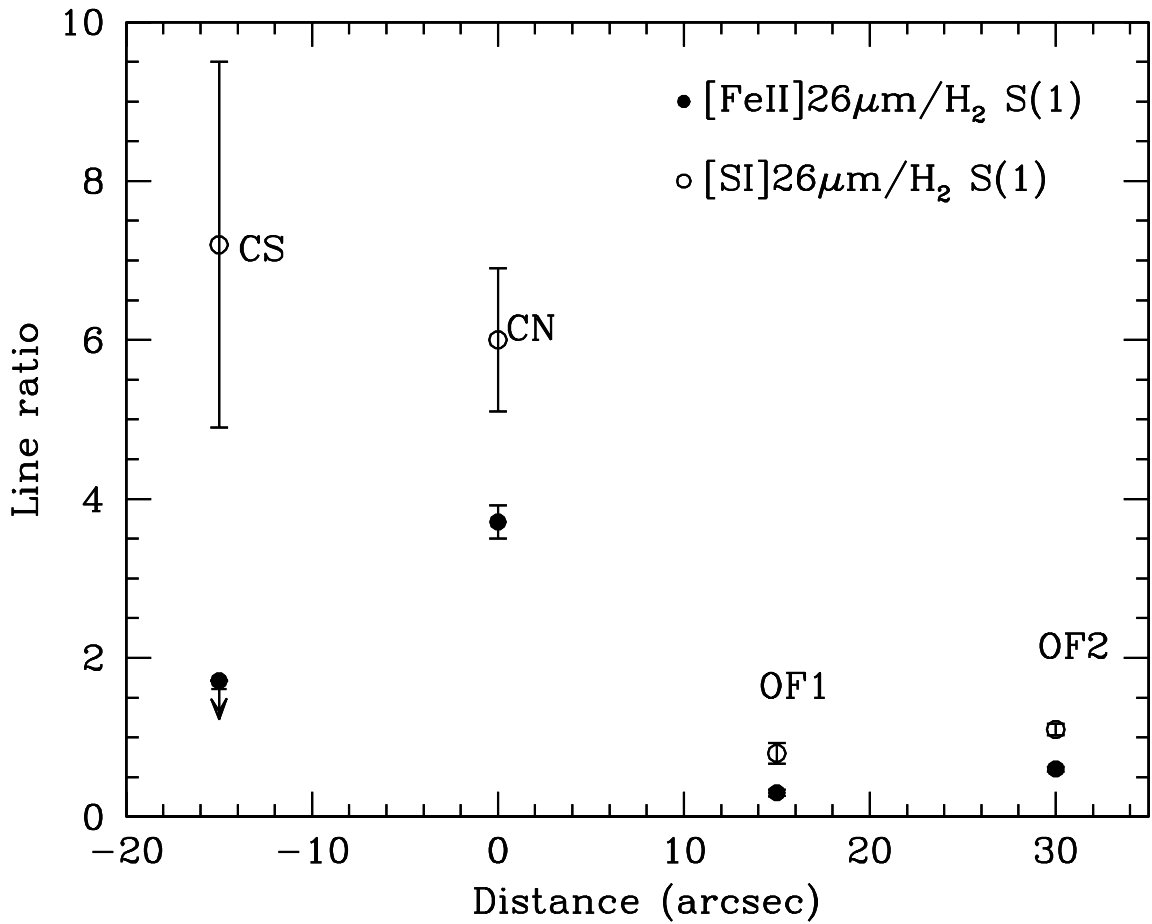}
\caption{Ratio of the atomic ([FeII]26$\mu$m and [SI]25$\mu$m)  over the molecular (H$_{2}$ S(1)) line emission. Atomic emission dominates close to the driving source (CN) while the atomic/molecular contribution decreases in the OF1 and OF2 positions (see Fig. \ref{fig1}).
\label{fig3}}
\end{figure}

Once extinction corrected, we have explored the spatial variation of the relative atomic/molecular emission, by plotting the ratio of the [FeII]26$\mu$m and [SI]25$\mu$m with respect to the H$_2$ 0--0 S(1) line (Figure \ref{fig3}). This plot clearly shows that the relative brightness of the atomic component with respect to the molecular component sharply decreases going from the CN  to the OF1 position while only slightly increasing again at the bow shock position. A similar behavior has been observed in the near-IR for a number of Class I jets, where the atomic and molecular components have been traced by the [FeII] 1.64$\mu$m and H$_2$ 1--0 S(1) lines respectively \citep{Bruni4, Bruni5}. 
  In these sources, the relative decrease of atomic gas emission with respect to H$_2$ in the jet beams is accompanied by a decrease of excitation in the jet which occurs on scales of $\sim$ 5-10$\arcsec$ from the driving source. The interpretation of this behavior is that the jet,  heated and ionized in the acceleration region, progressively expands and cools down until it strongly interacts with the medium through a bow shock.  The spatial scale of the IRS instrument is too poor to resolve the intensity and excitation variations within the jet expansion region; a qualitative inspection, however, suggests a similar behavior, in spite of the different physical conditions pertaining to the L1448 jet.

\subsection{$H_2$ emission}

Being very easily thermalized, H$_{2}$ lines can be used as probes for the temperature by means of excitation diagrams. The latter are constructed by plotting the values of \textit{ln(N$_{v,J}$/g)} against \textit{E$_{v,J}$}, where  N$_{v,J}$ is the column density for the population at the upper level, \textit{E$_{v,J}$} and \textit{g} are the level energy and statistical weight. Assuming LTE conditions, these two quantities are linearly related, and the local temperature can be derived from the slope of the fit. 
We have constructed excitation diagrams in each position combining the Spitzer and UKIRT observed lines. For these, column densities  have been derived dividing the line flux for the aperture adopted for the spectral extraction, assuming that the emission is extended and fills the extraction apertures in a uniform way.
This cannot be the case for the 0--0 S(0) line  observed with the much larger aperture of the LH module: the H$_2$ fundamental transition should be in fact also affected by diffuse emission from the cloud, therefore the derived column density can be considered as an upper limit of the S(0) column density in the jet.
In constructing the Boltzmann diagrams, we have assumed an H$_{2}$ ortho/para ratio equal to the equilibrium value of 3. 
We do not appreciate significant variations from this value inside the  errors of the data-points.

\begin{figure}
\epsscale{1.0}
\plotone{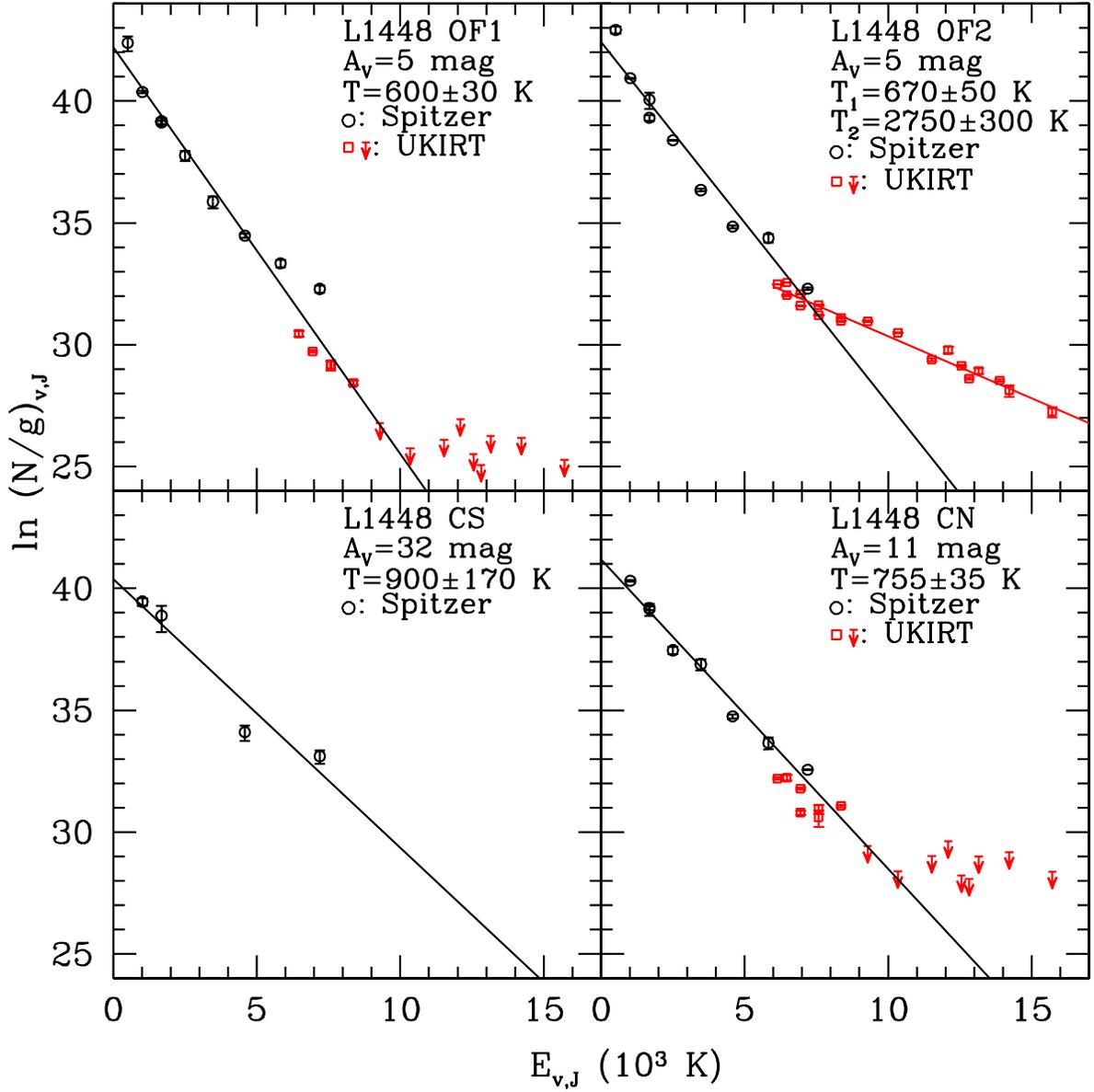}
\caption{Excitation diagram of the H$_2$ lines for each position defined in Fig. \ref{fig1}. Spitzer (open circles) and UKIRT (open squares and arrows for the upper limits) data points are optimally least square fitted (solid line). Temperatures and total column densities are derived from the slope of the fitted line, and its intersection with the \textit{ln(N$_{v,J}$/g)} axis. Visual extinction values used for dereddening and derived temperatures are displayed in the upper left corner of each panel.
\label{fig4}}
\end{figure}

The excitation diagrams for each position are presented in Figure \ref{fig4}; in these, Spitzer (open circles) and UKIRT lines (open squares) are least square fitted with a straight line. The visual extinction values used and temperatures derived from the slope of the fitted lines are displayed in the top right of each diagram. In all cases, data points are reasonably well aligned and thus giving support to the assumption of a H$_{2}$ gas in LTE. Because of the high values of visual extinction at CS, the lack of data points results in a fit with large uncertainties in the derived temperature. 

Given the fact that the lines falling within the Spitzer and UKIRT ranges originate from pure rotational and  ro-vibrational transitions, respectively, the two sets of lines may trace different temperature regimes. However it is only in the bowshock position  that  the two sets of lines clearly trace two different temperature components. In the other positions, 
the extinction corrected upper limits on high excitation H$_2$ lines suggest that a single temperature component with \textit{T} ranging between 600 
and 900 K, well represents the excitation conditions. 
It is only in the OF1 position that the column density of the S(6) and S(7) lines deviate from the single temperature fit: we think that the displacement of these lines is not reliable, since such curvature is not followed by the NIR measurements of the higher excitation lines.
From the Boltzmann diagrams it is also possible to derive the total H$_2$ column density of the warm gas, from  
 the intersection of the fitted line to the data points and the abscissa of the excitation diagram, which is equal to $ln N(H_2)/Q(T)$, with $Q(T)$ is the partition function at temperature $T$.

The derived temperatures and column densities for each of the observed regions are presented in Table \ref{tab3}. 
The gas temperature shows a slight decrease going from the driving source (CN) to the OF1 position, while column density displays a constant increase from the CS to the OF2 positions. This trend confirms that the excitation conditions decrease as the jet propagates into the ambient medium, 
and that in the inner region the jet is mostly atomic, as testified from the [FeII]/H$_2$ ratio displayed in Figure  \ref{fig3}.

\subsubsection{Shock model}

\begin{figure}
\epsscale{1.0}
\plotone{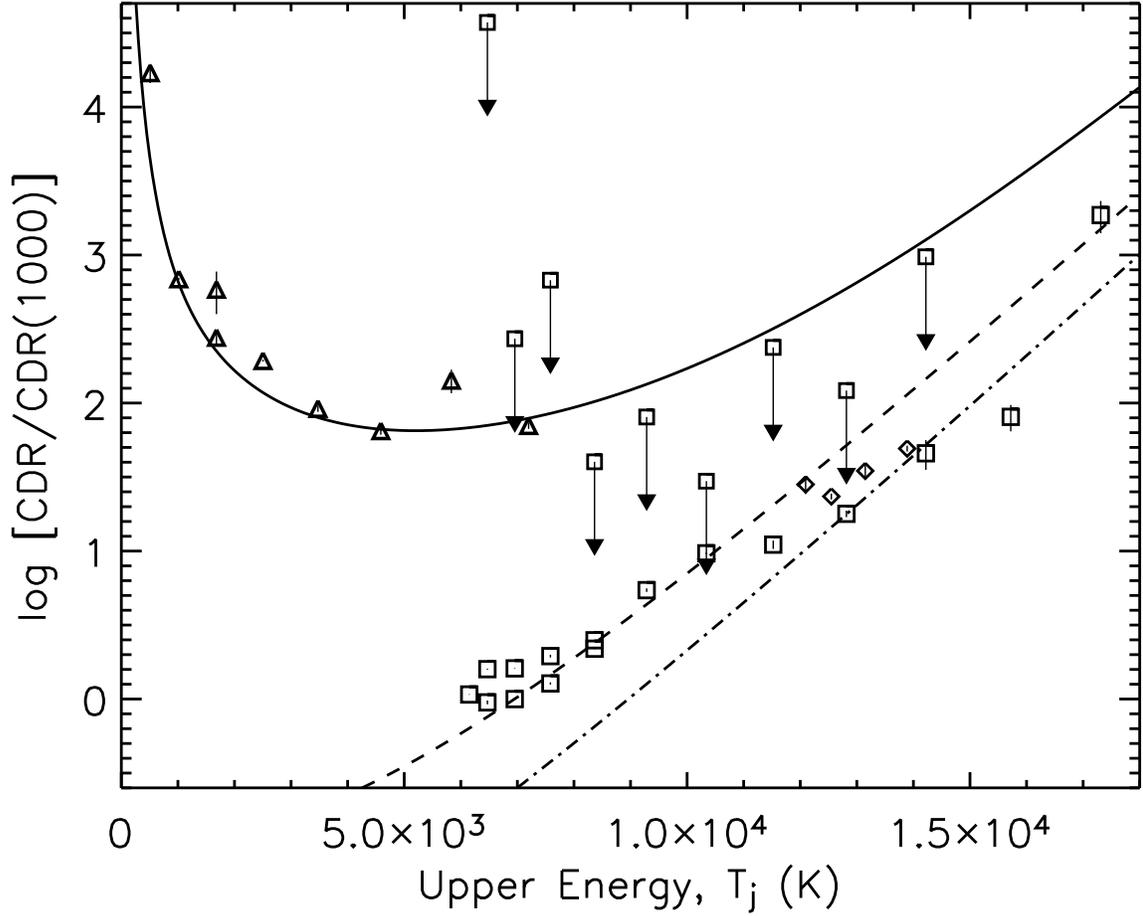}
\caption{Column Density Ratio (CDR) diagram of the OF2 position, constructed normalising  the column density of each transition 
 to the value given by a gas at 1000K and to the 1--0 S(1) column density.  Triangles represent Spizer 0-0 lines, squares represent lines 
from  the first vibrational levels, including the 1-1 upper limits derived from the Spitzer IRS spectra, and 
diamonds correspond to 2-1 S lines. The superimposed  model predictions are for the ground (solid), first (dashed) 
and second (dot-dashed line) vibrational level. The fit corresponds to a bow shock moving at 100~km~s$^{-1}$ into 
a medium of total density 10$^5$~cm$^{-3}$.
\label{fig5}}
\end{figure}

As seen in the previous section, the H$_2$ Boltzman diagram in the OF2 position
shows the presence of different temperature components, likely arising in the unresolved cooling zone  behind the bow shock.
In order to constrain the shock conditions from our Spitzer and UKIRT observations, we have attempted to model the derived 
 H$_2$ column densities with a C-type bow-shock model \citep{Smith}. 
For a better comparison of the derived column density with the model fit, we have constructed a Column Density Ratio (CDR) plot,
in which the column density in the upper energy level of each transition 
has been normalized to the value given by a gas at 1000K. These values are then presented 
relative to the column of the 1-0 S(1) upper energy of 6953K (see Fig.  \ref{fig5}). 
In this Figure, triangles represent Spitzer 0-0 lines, squares represent all lines originating 
from  the first vibrational levels, including the 1-1 upper limits derived from the Spitzer IRS spectra, and  
diamonds correspond to 2-1  lines; model predictions are superimposed over the data. 
The predicted column densities are calculated assuming  that the upper energy levels are in rotational LTE but 
vibrationally in non-LTE. This is a reasonable assumption for the levels at low excitation energy, while the higher rotational levels
appear to switch across from the 1-0 line to the 2-1 values, indicating that this assumption probably does not hold anymore for the higher levels.
The best fit model prediction is obtained with a  bow shock moving at 100~km~s$^{-1}$ into 
a medium of (H nucleon) density 10$^5$~cm$^{-3}$. The ion fraction of 10$^{-6}$ and magnetic field strength, 
corresponding to an Alfven speed of 2 km~s$^{-1}$, imply that the non-dissociative wings of the bow are C-type shocks.
The bow geometry required for the obtained fit is such that the bow must possess a compact apex and extended 
flanks with a shape $Z \propto R^s$ with $s = 1.38$ in cylindrical coordinates. Then, a large column of low 
excitation gas is generated by the bow.  
A C-bow shock model with similar parameters as adopted here was able to reproduce also the kinematic and the 1-0 S(1)  
morphology of the considered bow shock \citep{davis}.

\subsection{Atomic Jet \label{sub1}}

As pointed out in Section \ref{SPITZER}, the detection of only the fundamental [FeII] transition indicates that the excitation conditions of the atomic jet are very low. 
The [FeII]~26$\,\mu$m line originates from the $^6D_{7/2}$ level, that has an excitation temperature of $\sim$550 K, while the level just above this one,  $^6D_{5/2}$, gives origin to a transition at ~35$\,\mu$m, having $T_{ex} \sim$ 960 K. 
In principle, the non detection of the 35~$\,\mu$m transition could give a strong constraint on the gas temperature. However, the spectral region around 30~$\,\mu$m is very noisy and the upper limits are not stringent. 
To get constraints on the temperature, we have instead considered  the upper limit on the $^4F_{7/2}-^4F_{9/2}$ transition at ~18$\,\mu$m, that lies in a region of lower noise. In addition,  we have also used  the ratio [SiII]35$\,\mu$m/[FeII]26 $\,\mu$m as a density probe.
Si and Fe have a comparable ionization potential (8.15 and 7.9  eV respectively)  thus a similar degree of ionization. Furthermore these two lines are excited at a similar temperature of $\sim$ 500 K, thus their ratio depends only on the electron density and on the [Si/Fe] gas phase abundance ratio.
We assume a [Si/Fe]$_{gas}$ ratio equal to the solar ratio \citep[taken from][]{GA}, an assumption implying that the two species are equally depleted on grains. The validity of this assumption is further examined in Section \ref{refractory}.

\begin{figure}
\epsscale{1.0}
\plotone{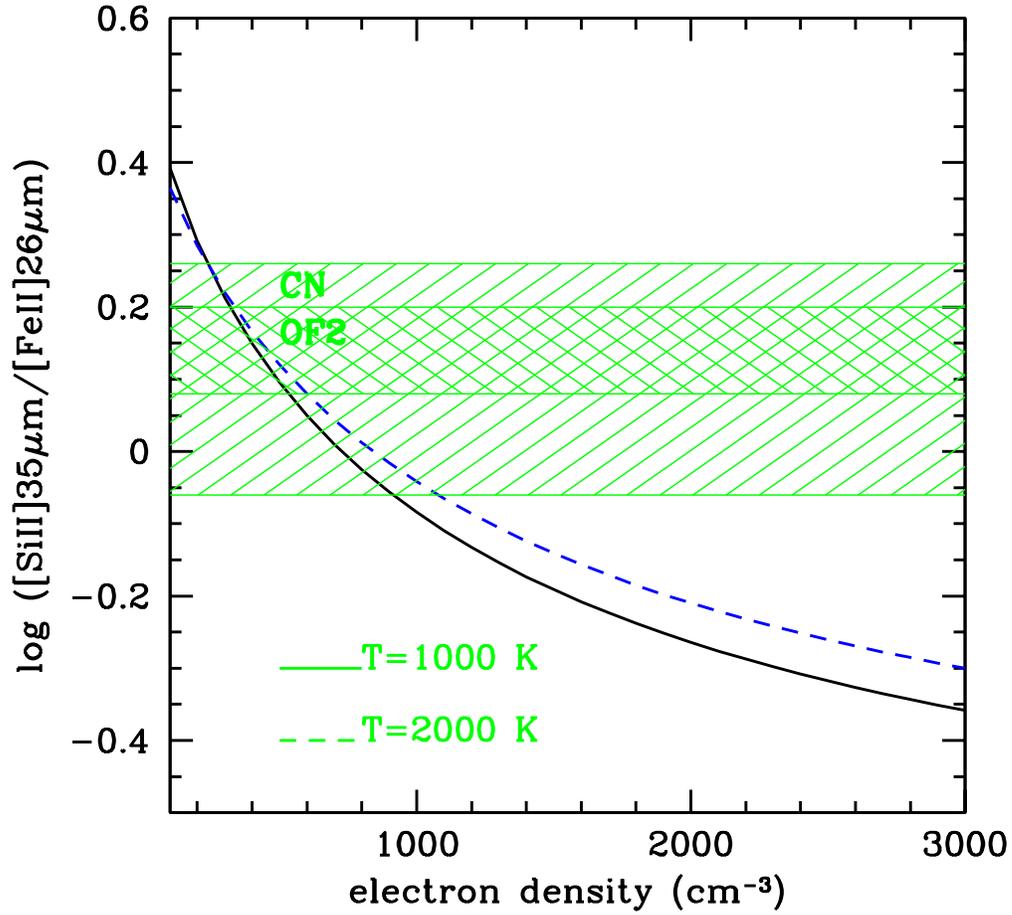}
\caption{Diagnostic diagram of the [SiII]35$\mu$m/[FeII]26$\mu$m ratio versus electron density, for temperatures of 1000K and 2000K (solid and dashed lines). Hatched areas represent the observed ratio for the CN and OF2 regions that correspond to electron densities of from 200  to 1000 cm$^{-3}$ and from 300 to 500 cm$^{-3}$ respectively. 
\label{fig6}}
\end{figure}

Figure  \ref{fig6} presents a plot of the [SiII]34.8$\mu$m/[FeII]26.0$\mu$m ratio as a function of the electron density, for temperatures of 1000K and 2000K (solid and dashed lines). This diagram has been constructed employing a statistical equilibrium model that considers the first 16 levels for [FeII] \citep{Bruni4}, and a two level system for [SiII]. Radiative and collisional rates for the [SiII] have been taken from \citet{dufton}.
Hatched areas represent the  ratios observed in the CN and OF2 regions,  respectively. Such observations are consistent with values of n$_e$ in the range $\sim$300-500 for the OF2 position and $\sim$200-1000 cm$^{-3}$ for the CN position. 

\begin{figure}
\epsscale{1.0}
\plotone{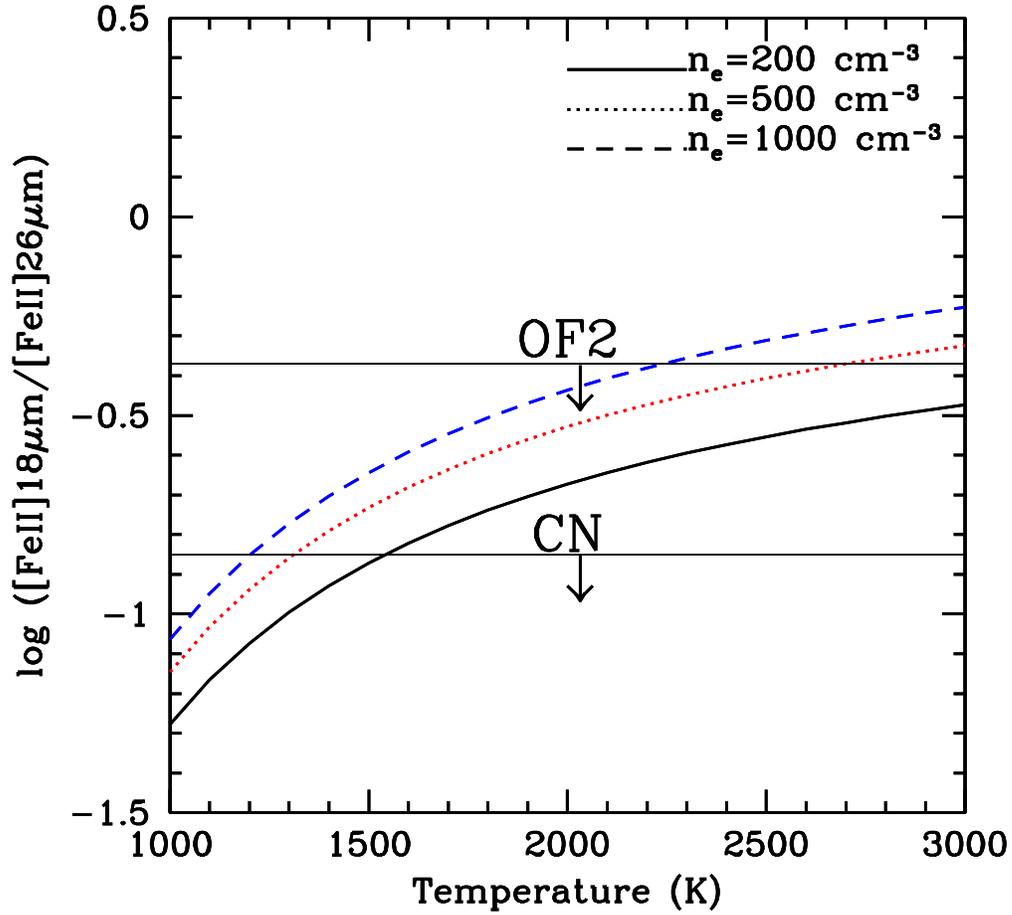}
\caption{Diagnostic diagram of [FeII]18$\mu$m/[FeII]26$\mu$m ratio versus temperature, for electron densities of 2 10$^2$, 5 10$^2$ and 10$^3$cm$^{-3}$ (solid and dashed lines respectively). The non-detection of the[FeII]18$\mu$m line allows us only to constrain the upper temperature limits for the CN and OF2 positions
\label{fig7}}
\end{figure}

In Figure  \ref{fig7}, the upper limit ratio  [FeII] 18$\mu$m to 26$\mu$m is plotted as a function of the gas temperature, for different electron density values. The 26$\mu$m line flux and the 18$\mu$m line upper limit have been measured on  two IRS modules having different field of views and therefore their intrinsic ratio depends upon the extension of the emitting region. A conservative upper limit on this ratio is obtained assuming beam filling and consequently normalizing the line ratio to the different field of view. The upper limits obtained in this way 
are displayed in Fig.  \ref{fig7} for the CN and OF2 regions. 
From this plot, we derive that gas temperatures less than $\sim$ 2500 K and $\sim$ 1500 K are responsible for the emission observed in the OF2 and CN positions, respectively.   

The derived  physical conditions significantly differ from the conditions measured from [FeII] 
near-IR lines in jets from more evolved class I sources, that have temperatures ranging from 7000 and 15000 K and densities $\sim$ 10$^{4}$ -10$^{5}$ cm$^{-3}$ \citep[e.g.][]{Bruni4, Takami} . Low electron density values, in particular, point to a low ionization fraction for the gas under consideration.
Total n$_{H2}$ densities of the order of 10$^5$ cm$^{-3}$ or higher have been inferred in the L1448 jet from sub-mm and far-IR observations \citep{Bruni2, Bruni6}, implying a ionization fraction x$_e$ of  $\sim$ 10$^{-2}$ or lower. 

The inferred low temperature and ionization fraction, in conjunction with the non-detection of the [NeII] 12.8 $\mu$m line,  imply that, if these lines are shock excited, the shock velocity should be low. Fast $J$-type shocks (e.g. Hollenbach \& Mc Kee 1989) are therefore excluded. 
Low-velocity J-type shocks might provide sufficient ionization to excite the fundamental ionic fine-structure lines, but the shock velocity needs to be lower  than $\sim$ 10 km\,s$^{-1}$ to have a temperature of less than $\sim$2000 K.

Part or all of the atomic emission observed on-source may in principle also originate from excitation in a circumstellar disk. Line emission disk models (Gorti \& Hollenbach 2008) as well as Spitzer observations in T Tauri disks (e.g. Pascucci et al. 2007, Lahuis et al. 2007) show however that disk emission is characterized by both strong [NeII] 12.8 $\mu$m, excited in the high temperature gas heated and ionized 
at the disk surface, and  [FeI] 24 $\mu$m line, originating in the deeper disk vertical layers, that we do not detect here.

The strong atomic emission observed on-source may instead originate from the jet base, in a zone similar to the Forbidden Emission Line (FEL) regions observed in Class I/II jets although with very different excitation conditions \citep[e.g.][]{davis03}.  In analogy with the FEL regions there is a decrease in intensity of the atomic emission with distance from the source, likely caused by the expansion of the jet and the consequent cooling down of the gas. 

\section{Dust disruption and abundance of refractory species \label{refractory}}

Gas phase abundances of Iron and Silicon are very low in the interstellar medium since these refractory elements are easily depleted onto the cores of dust grains. Shocks occurring along the outflows of young stars are able to at least partially restore 
the refractory elements to the gas phase, through processes like sputtering and grain-grain collisions \citep{jones}. Observations of mm and sub-mm  SiO  lines in many molecular outflows \citep[e.g.][]{claudio, gibb, Bruni6} have shown that dust grains are indeed partially destroyed, and that the released Si undergoes chemical reactions leading to the formation of SiO. \citep{schilke, antoine}. SiO abundance determinations along the molecular jet of L1448 show however that gas-phase Si locked in SiO is only about 5\,$10^{-3}$ of the Si solar abundance: thus either not all of the Si released by the dust reprocessing is converted into SiO, or the shocks are not able to completely restore all the Si to the gas phase. The detection of the [SiII] fundamental line in our spectra suggests that indeed a significant part of gas-phase Si is present in 
ionic form.
On the other hand, the gas-phase abundance of Fe in jets have been so far measured only in near-IR jets of Class I sources, from the bright near-IR [FeII] lines \citep{Bruni4, Bruni5, linda}. Such studies have shown that a large fraction of Fe (from 70 to 95\%) is still locked in grains, indicating that dust grains have not been totally destroyed by shocks. No estimates have been so far given about the iron gas-phase abundance in Class 0 molecular outflows.
We can now provide abundance estimates of both Si and Fe through the detected emission lines of their single-ionized atoms.

The gas phase abundance of refractory elements can be derived from a comparison of their emission lines with those of a non-refractory species emitted under the same physical conditions. In the case of our spectra,  the [SI]25.2$\mu$m line can be used as a reference, as sulfur is not depleted in grains, assuming that all the sulfur is in neutral form.  To check  this hypothesis, we have computed the S$^0$/S$^{+}$ ratio applying ionization equilibrium between collisional and charge-exchange ionization, and direct/dielectronic 
recombination \citep[rates from][]{stancil, Landini}. For the inferred physical conditions of 
$T\sim$ 1000-2000 K and x$_e \la$ 10$^{-2}$, it results that 95$\%$ of S is in neutral form. 
The iron gas phase abundance, relative to the solar value,  can be therefore written as:

\begin{equation}
[Fe]_{gas}/[Fe]_{\odot} = \frac{[Fe/S]}{[Fe/S]_{\odot}}=\frac{F([FeII]26\mu m)}{F([SI]25.2\mu m)}\times \frac{\epsilon([SI]25.2\mu m)}{\epsilon( [FeII]26\mu m)}\times [S/Fe]_{\odot}
\end{equation}

where $\epsilon([SI]25.2\mu m)$ and $\epsilon( [FeII]26\mu m)$ are the theoretical emissivities, $[Fe/S]_{\odot}$ is the solar abundance ratio, taken from  \citet{asplund}, and F([FeII]26$\mu$m)/F([SI]25.2$\mu$m) is the observed ratio. A similar expression can be written for the $[Si]_{gas}/[Si]_{\odot}$ ratio.
To compute the theoretical emissivities, we have considered for S a 5 level statistical equilibrium code assuming electronic collisional excitation and  n$_e$=400 cm$^{-3}$. Two values of temperatures have been considered: 2500 K (the upper limit derived in the OF2 position) and 600 K (the lower value derived by the H$_2$ analysis).
In a medium with high total density and low ionization fraction, such as the one we are considering here, collisions with atomic hydrogen may become important in the excitation of atomic species such as [SI], in spite of the low  n$_H$ collisional de-excitation rates with respect to the  rates for electronic collisions ($\gamma_{H}/\gamma_{e} \sim 10^{-4}$ for the [SI]25$\mu$m line, \citet{hollenbach}). In order to asses how collisions with hydrogen affect the results, we have calculated the emissivities also assuming a medium with a total density of 10$^5$ cm$^{-3}$ and n$_H$ = 0.1 n$_{tot}$ \footnote{neutral hydrogen collisional rates for ions, such as [FeII] and [SiII], are about 10$^6$ times weaker than electron collisional rates: we have checked that n$_H$ collisions start to give comparable ion emissivities with respect to electron collisions  only for n$_H > 10^5$ cm$^{-3}$ and n$_e < 500$ cm$^{-3}$. We have therefore not considered collisions with n$_H$ in our calculations of [FeII] and [SiII] emissivities } .

We have applied the above analysis to the values observed in the CN, OF1 and OF2 positions: for the [SI]25.2$\mu$m flux, we have taken the average value given by the determinations obtained in the two different orders of the LH module. Results are summarized in Table \ref{tab4}: the uncertainty in temperature results in a factor of two the uncertainty in derived abundances, while a difference of four is found between results obtained assuming collisions with electrons and with atomic hydrogen.  

The Table shows that the gas phase abundance of Fe  and Si remains between 5 and 20\% of the solar values for both species and in all positions. Similar low values of Fe abundances have been also found in the inner regions of  jets from Class I sources, from the analysis of the [FeII] NIR lines \citep{Bruni5, linda}, while larger abundances (from 20 to 70\% of the solar value) have been derived at large distances from the driving source and in bow shocks \citep{Bruni4, teresa}.  

Finally, taking a SiO abundance of 5\,$10^{-3}$ derived by \citet{antoine}, the Si/SiO abundance ratio is $\sim$ 100, giving support to the hypothesis  that most of the silicon released from grains remains in atomic form. 

\section{Mass and mass flux in the warm gas}

The presented observations point to the presence of warm gas associated with the L1448 CO and SiO millimeter jet, composed of both a molecular and a weakly ionized component. We can assess if this warm gas represents a dynamically important component of the jet, by measuring 
its mass flux  and comparing it with estimates of the mass flux based on ISO and sub-mm observations of the CO emission \citep{Bruni2, bachiller1}.   
We can determine the mass flux for both the molecular and atomic counterparts of the outflow, using as tracers H$_{2}$ and [FeII]/[SI]  respectively. The method is similar in both cases and is based on the fact that the line emission  is optically thin, so that the observed luminosity is proportional to the mass of the emitting gas \citep{hartigan}. For [FeII] and [SI] we have applied the relationship given in \citet{Bruni5}: 

\begin{equation}
\dot{M} = \mu m_{H}\times(n_{H}V)\times(dv_t/dl_t)
\end{equation}

where $\mu$ is the mean atomic weight, m$_H$ is the proton mass, n$_H$  is the total number density, V the volume of the emitting region, dl$_t$ is the projected length perpendicular to the line of sight and dv$_t$ the tangential velocity of the observed region. The number of emitting atoms can be derived from the line luminosity according to the relation:

\begin{equation}
n_{H}\,V=L(line) \left(h\nu A_i f_i \left[\frac{X}{H}\right] \right )^{-1}
\end{equation}

where \textit{A$_i$} and \textit{f$_i$} are the radiative rate and fractional population of the upper level,  and [X/H] is the gas phase  abundance of the considered atom/ion. For the determination of the mass flux from the [FeII]26$\mu$m line, we have assumed that all iron is singly ionized based on the
fact that no neutral iron lines are observed within the IRS range; this assumption is based on the fact that the [Fe I] ground transition at 24$\mu$m is not detected in any of our spectra, in spite of its high radiative rate coefficient, comparable to that of the [FeII] ground transition. We have taken the iron gas phase abundance estimated in Sect 3.1 and listed in Table \ref{tab4}. 
Conversely, to apply the relationship (3) to the [SI]25$\mu$m line, we have assumed that sulfur is all neutral (as discussed in Sect. 3.1) and taken the sulfur solar abundance.

The  projected length dl$_t$ was taken equal to the jet length sampled by the width of the LH slit. 
Finally, the tangential velocity of the outflow was taken equal to 170 km s$^{-1}$ as derived from the SiO proper motion study by \citet{GA}.
Such a velocity, appropriate for the sub-mm jet, may be just a lower limit for the atomic jet velocity, if the latter is associated with a higher velocity inner component. It has been shown however that atomic jets from Class I/II sources have velocities in the range
100-300 km\,s$^{-1}$ (Ray et al. 2007, Davis et al. 2003). We therefore believe that our assumption can introduce an uncertainty of at most a factor of two in the mass flux determination.
The derived mass flux values using the above method are listed in Table \ref{tab5}. The uncertainty associated with the Fe gas-phase abundance 
results in  an order of magnitude uncertainty in the mass flux determination using the [FeII]26$\mu$m. The determinations from the [SI] line agree with the range of values derived from [FeII] and  indicate a mass flux of the order of 1-2\,10$^{-6}$ $M_{\sun}$ yr$^{-1}$  in the CN position, while lower values, of the order of 3-8\,10$^{-7}$ $M_{\sun}$ yr$^{-1}$, are estimated at the bow-shock OF2 position.

The determination of the mass flux from the H$_2$ component has been measured instead from the total column density for each extracted region as derived from the excitation diagrams, applying the relationship:
\begin{equation}
\dot{M} = \mu m_H\times(2\,N(H_2)A)\times(dv_t/dl_t)
\end{equation}

where N(H$_2$) is the total column density and A is the area sampled by the slit. We consider here again the tangential velocity as measured from the SiO proper motion, assuming that it is representative of all the molecular gas. The derived mass flux along the outflow with this method is presented in Table \ref{tab5}. The H$_2$ mass flux in the central position is about two orders of magnitude lower than the value derived from the atomic lines  while in the OF2 position \.{M}(H$_2$) is a factor between 2 and 5 lower than \.{M}([SI]).Thus the bulk of the flowing mass is carried out by the atomic jet all along the sampled positions.

A mass flux of the order of   5\,10$^{-6}$ $M_{\sun}$\,yr$^{-1}$ has been derived within 20$^{\arcsec}$ of the driving source combining multi-transition CO sub-mm and ISO data \citep{Bruni2}. This value is only a factor of few  larger than the value we derive from atomic lines at the CN position, which suggests that the atomic jet detected by our observations and the warm CO outflows sampled by the ISO and sub-mm observations are dynamically linked.
It is also instructive to examine if the atomic jet detected by our Spitzer observations posseses enough momentum to sustain the entrained CO outflow observed at mm
wavelength. \citet{bachiller1} derived, for the L1448-mm blue-shifted lobe, a momentum flux of the order of 2.6\,10$^{-4}$ M$_{\sun}$\,km\,s$^{-1}$\,yr$^{-1}$, assuming an outflow inclination of 70$^o$ with respect to the plane of the sky. If we assume momentum conservation for entrainment of the outflow by the jet, and a total jet velocity of 170 km\,s$^{-1}$, we derive that the primary jet should possess a mass  flux of the order of 1.4\,10$^{-6}$ $M_{\sun}$\,yr$^{-1}$ . This is consistent with the values we infer from the atomic emission, supporting the hypothesis that  the detected atomic jet represents the energetically most important component of the outflow.

The accretion rate can be also estimated assuming the bolometric luminosity is mainly due to accretion. Using a stellar mass $M_\star$ $\sim$ 0.5 M$_\odot$ \citep{froebrich2} and a stellar radius $R_\star$ $\sim$ 4 R$_\odot$ \citep{stahler}, we get an accretion rate \.M$_{acc} \sim L_{bol}\,R_\star/G\,M_\star \sim$ 2\,10$^{-6}$ $M_{\sun}$\,yr$^{-1}$. 
We thus derive a value comparable to the mass ejection rate estimated above. This is inconsistent with the picture that about only 10\% of the mass accreting on the protostar is ejected by means of magneto-centrifugal acceleration mechanisms \citep[e.g.][]{ferreira}. This may indicate that the actual protostellar mass is lower than assumed, and that the source is still at the beginning of its main accretion phase. 

\section{Conclusions \label{conclusions}}

We have carried out Spitzer IRS spectroscopic observations towards the molecular jet driven by the Class 0 source L1448-C. We have detected the  H$_2$ pure rotational lines from S(0) to S(7) alongside fundamental fine-structure transitions of Si$^+$, S$^0$ and Fe$^+$. Additional UKIRT/UIST spectra have been acquired, showing that, at variance with the 0--0 lines,  near-IR H$_2$ vibrational line emission is mainly confined to the bow shock region where the jet and the ambient molecular cloud interact. 
We have analysed these lines in order to derive the physical and dynamical parameters of the associated warm gas. The main results can be summarised as follow:

\begin{itemize}
\item The detection of the fine-structure lines testify for the presence to a previously unknown atomic jet, embedded in the molecular outflow. 
An analysis based on the line ratios as well as on previous observations at sub-mm wavelengths indicates that the excitation conditions of this jet are very low. In the region close to the driving source, a temperature T$_e <$ 1500 K and an electron density n$_e \sim$200-1000 cm$^{-3}$  have been measured, while the estimated ionization fraction is surprisingly $<$ 10$^{-2}$. 
We suggest that the detected atomic gas represents the analog of the  Forbidden Emission Line (FEL) region observed in the more evolved Class I/II sources, although with very different excitation conditions. The close similarity with the FEL region is seen in the decrease of intensity of the atomic emission with distance from the source; in both cases this is likely caused by the expansion of the jet and the consequent gas cool down.  Observations with better spatial and spectral resolution, like those that will be possible with the Mid-Infrared Instrument (MIRI) instrument on the James Webb Space Telescope (JWST), are needed to have a better picture of the origin of this gas through comparison with jet acceleration models. 

\item The H$_2$ rotational emission indicates the presence of warm gas at a temperature ranging between 600 and 900 K.  It is only at the bow shock position that a second component at higher temperature is detected in the near-IR. The H$_2$ emission  in the inner 
jet region may represent warm molecular gas enveloping the atomic gas: such a molecular component can be also related with the 
collimated SiO jet observed at sub-mm wavelengths, the excitation conditions of which are similar to those inferred here for the H$_2$. 

At the bow shock position, the near and mid-IR H$_2$ line observations have been compared with a C-type bow shock model. 
Shock velocities of $\sim$ 100 km\,s$^{-1}$ and pre-shock densities n(H$_2$) $\sim$ 10$^5$ cm$^{-3}$ can account rather well for the observed column density.

\item The gas phase abundance of Fe and Si has been estimated to infer the amount of dust reprocessing in the jet. Only a fraction of $\sim$0.1-0.3 of the solar abundance of  Fe and Si is in gaseous form indicating that dust still plays a major role in the depletion of refractory elements. A comparison with the SiO abundance recently derived from the analysis of sub-mm transitions shows that the Si/SiO abundance ratio is $\sim$100, and thus that most of the silicon released from grains by sputtering and grain-grain collisions remains in atomic form. The inferred Fe gas-phase abundance is similar to the values previously estimated in Class I jets by means of near-IR lines: this suggests  that dust reprocessing does not depend on evolutionary status of the outflows although further studies are required to confirm this.

\item Estimates of the atomic and molecular mass flux rates have been derived from the luminosity of the outflows and the kinematical information given by sub-mm interferometric data. Values of the order of $\sim$ 10$^{-6}$ and $\sim$10$^{-7}$ M$_{\sun}$\,yr$^{-1}$ are inferred from the [SI]25$\mu$m and  H$_2$ line luminosities, respectively. A comparison with the momentum flux of the large scale CO molecular outflow suggests that the detected atomic jet has the power to drive the large scale outflow and thus may represent the primary jet ejected by the source.

\end{itemize}

\acknowledgments
The present work was supported in part by the European Community’s
Marie Curie Actions - Human Resource and Mobility within the JETSET (Jet
Simulations, Experiments and Theory) network under contract MRTN-CT-2004
005592.
This work is based on observations made with the Spitzer Space Telescope, which is operated by the Jet Propulsion Laboratory, California Institute of Technology under a contract with NASA, and with the the United Kingdom Infrared Telescope, operated by the Joint Astronomy Centre on behalf of the Science and Technology Facilities Council of the U.K.

\clearpage
%
%
%
%
%
%
\input{tab1.tex}
%
\input{tab2.tex}
%
\input{tab3.tex}
%
%
\input{tab4.tex}

\input{tab5.tex}
%
\clearpage

\end{document}

%% file: tab1.tex
\begin{deluxetable}{ccccccc}
\tablecaption{L1448 observed lines:SPITZER/IRS \label{tab1}}
\tablecolumns{7}
\tablewidth{0pt}
\tablehead {\multicolumn{3}{c}{Lines} & \multicolumn{4}{c}{$F\pm\Delta~F$(10$^{-14}$erg\,cm$^{-2}$\,s$^{-1}$)}\\
\hline\\[-5pt]
\colhead{$\lambda$($\mu$m)}  &  \colhead{Element(Transition)} & \colhead{Module}  & \colhead{CS} & \colhead{CN} & \colhead{OF1} & \colhead{OF2}}
\startdata
5.51116   &    H$_2$  0-0   S(7)    &  SL  &   5.5$\pm$1.5  &    4.5$\pm$0.1   &    3.8$\pm$0.6 &       3.9$\pm$0.2  \\
6.10856   &    H$_2$  0-0   S(6)    &  SL  &   $<$ 0.7       &    2.1$\pm$0.5   &    1.7$\pm$0.3 &       4.8$\pm$0.9  \\
6.90952   &    H$_2$  0-0   S(5)    &  SL  &   3.2$\pm$1.0  &    7.8$\pm$0.6   &    6.4$\pm$0.6 &       9.3$\pm$0.5  \\
8.02505   &    H$_2$  0-0   S(4)    &  SL  &   $<$ 2       &    7.0$\pm$1.5   &    2.8$\pm$0.7 &       4.5$\pm$0.2  \\
9.66491   &    H$_2$  0-0   S(3)    &  SL  &   $<$ 0.5       &    4.9$\pm$0.7   &    10.8$\pm$2.2&       20.1$\pm$0.3 \\
12.2786   &    H$_2$  0-0   S(2)    &  SL  &   1.1$\pm$0.4  &    2.8$\pm$0.7   &    3.5$\pm$0.4 &       4.0$\pm$0.4  \\
12.2786   &    H$_2$  0-0   S(2)    &  SH  &   1.7$\pm$0.2  &    2.9$\pm$0.2   &    3.3$\pm$0.2 &       8.4$\pm$0.3  \\
17.0348   &    H$_2$  0-0   S(1)    &  SH  &   0.4$\pm$0.1  &    2.4$\pm$0.1   &    3.2$\pm$0.2 &       5.7$\pm$0.1  \\
28.2188   &    H$_2$  0-0   S(0)    &  LH  &  $<$ 0.6       &       $<$ 1 &    1.2$\pm$0.4 &       2.1$\pm$0.3  \\
\hline\\ [-10pt]
 5.33042  &    H$_2$  1-1   S(8)    &  SL  &  $<$ 0.8   & $<$ 2       & $<$ 1.3  &  $<$ 2     \\
 5.81112  &    H$_2$  1-1   S(7)    &  SL  &  $<$ 0.8   & $<$ 1       & $<$ 1    &  $<$ 1.6   \\
 6.43710  &    H$_2$  1-1   S(6)    &  SL  &  $<$ 1.3   & $<$ 1.7     & $<$ 1.2  &  $<$ 1.8   \\
 7.28070  &    H$_2$  1-1   S(5)    &  SL  &  $<$ 2     & $<$ 1.6     & $<$ 1.5  &  $<$ 0.9   \\
 8.45367  &    H$_2$  1-1   S(4)    &  SL  &  $<$ 1.2   & $<$ 0.8     & $<$ 0.5  &  $<$ 0.8   \\
 10.1777  &    H$_2$  1-1   S(3)    &  SL  &  $<$ 0.5   & $<$ 0.8     & $<$ 0.9  &  $<$ 0.8   \\
 12.9275  &    H$_2$  1-1   S(2)    &  SL  &  $<$ 0.8   & $<$ 0.5     & $<$ 0.8  &  $<$ 1.8   \\
 17.9320  &    H$_2$  1-1   S(1)    &  SH  &  $<$ 0.5   & $<$ 0.6     & $<$ 0.4  &  $<$ 0.4   \\
 29.7017  &    H$_2$  1-1   S(0)    &  LH  &  $<$ 2     & $<$ 2       & $<$ 0.9  &  $<$ 0.8   \\
\hline\\ [-10pt]
5.34017   &   [FeII] $^{4}$F$_{9/2}$ -- $^{6}$D$_{9/2}$  &  SL  &   $<$ 2           &   $<$ 1          &   $<$ 2        &       $<$ 3           \cr
17.9359   &    [FeII] $^{4}$F$_{7/2}$ -- $^{4}$F$_{9/2}$ &  SH  &   $<$ 0.4         &   $<$ 0.3        &   $<$ 0.5      &      $<$ 0.4           \cr
25.9883   &    [FeII] $^{6}$D$_{7/2}$ -- $^{6}$D$_{9/2}$ & LH   &   $<$ 1           &    9.7$\pm$0.4   &  1.2$\pm$0.1   &       4.2$\pm$0.1  \cr
35.7774   &    [FeII] $^{4}$F$_{3/2}$ -- $^{4}$F$_{5/2}$ & LH   &   $<$ 2           &   $<$ 3          &   $<$ 3        &      $<$ 1.5       \cr
\hline\\[-10pt]
25.2490\tablenotemark{a}   &    [SI]  $^{3}$P$_{1}$ -- $^{3}$P$_{2}$       &  LH  &   5.8$\pm$0.3  &    15.2$\pm$0.5  &    3.2$\pm$0.2 &       6.8$\pm$0.2  \cr
25.2490      &    [SI]  $^{3}$P$_{1}$ -- $^{3}$P$_{2}$       &  LH  &   5.3$\pm$0.4  &    19.2$\pm$0.5  &    2.6$\pm$1.2 &       6.1$\pm$0.2  \cr
34.8152   &    [SiII] $^{2}$P$_{3/2}$ -- $^{2}$P$_{1/2}$   & LH   &   $<$ 8  &    12.9$\pm$4.2 &    $<$ 3      &       5.6$\pm$0.6 
\enddata
\tablenotetext{a}{the [SI]25.2$\mu$m line is detected in two different orders of the LH module.}
\end{deluxetable}

%% file: tab2.tex
\begin{deluxetable}{ccccc}
\tablecaption{L1448 observed  lines:UKIRT/UIST \label{tab2}}
\tablewidth{0pt}
\tablehead {\multicolumn{2}{c}{Lines} & \multicolumn{3}{c}{$F\pm\Delta~F$(10$^{-16}$erg\,cm$^{-2}$\,s$^{-1}$)}\\
\hline\\[-5pt]
\colhead{$\lambda$($\mu$m)}  &  \colhead{Element(Transition)} & \colhead{CN} & \colhead{OF1} & \colhead{OF2}}
\startdata
1.6877 & H$_2$ 1-0 S (9)                 &  $<$ 0.3            &   $<$ 0.04              & 0.15$\pm$0.03 \\
1.7147 & H$_2$ 1-0 S (8)                &  $<$ 0.4             &   $<$ 0.05              & 0.17$\pm$0.04 \\
1.7480 & H$_2$ 1-0 S (7)                 &  $<$ 0.4            &   $<$ 0.05              & 0.94$\pm$0.05 \\
1.7880 & H$_2$ 1-0 S (6)                 &  $<$ 0.4            &   $<$ 0.05              & 0.75$\pm$0.05 \\
1.9576 & H$_2$ 1-0 S (3)                 &  $<$0.3             &   0.52$\pm$0.05         & 8.10$\pm$0.04 \\
2.0338 & H$_2$ 1-0 S (2)                 &  1.36$\pm$0.45        & 0.29$\pm$0.04           & 2.65$\pm$0.05 \\
2.0735 & H$_2$ 2-1 S (3)                 &  $<$ 0.5             &   $<$ 0.06              & 1.00$\pm$0.05 \\
2.1218 & H$_2$ 1-0 S (1)                 &  3.51$\pm$0.33        & 1.00$\pm$0.05           & 7.90$\pm$0.04 \\
2.1542 & H$_2$ 2-1 S (2)                 &  $<$ 0.4             &   $<$ 0.05              & 0.38$\pm$0.05 \\
2.2235 & H$_2$ 1-0 S (0)                 &  $<$ 0.4             & 0.35$\pm$0.05           & 2.10$\pm$0.06 \\
2.2477 & H$_2$ 2-1 S (1)                &  $<$ 0.4             &   $<$ 0.05              & 0.97$\pm$0.04 \\
2.3556 & H$_2$ 2-1 S (0)                 &  $<$ 0.3             &   $<$ 0.04              & 0.32$\pm$0.04 \\
2.4066 & H$_2$ 1-0 Q (1)                 &8.07$\pm$0.39          & 1.31$\pm$0.04           & 9.73$\pm$0.04 \\
2.4134 & H$_2$ 1-0 Q (2)                 &3.28$\pm$0.36          & 0.45$\pm$0.05           & 4.10$\pm$0.03 \\
2.4237 & H$_2$ 1-0 Q (3)                 &8.01$\pm$0.37          & 1.56$\pm$0.05           & 9.86$\pm$0.04 \\
2.4375 & H$_2$ 1-0 Q (4)                 &1.49$\pm$0.28          & 0.42$\pm$0.05           & 2.59$\pm$0.06 \\
2.4548 & H$_2$ 1-0 Q (5)                 &5.89$\pm$0.30          & 1.09$\pm$0.05           & 5.34$\pm$0.04 \\
2.4756 & H$_2$ 1-0 Q (6)                &  $<$ 0.4             &   $<$ 0.05              & 1.75$\pm$0.05 \\
2.5001 & H$_2$ 1-0 Q (7)                 &  $<$ 0.5             &   $<$ 0.06              & 3.56$\pm$0.06 \\
\enddata
\end{deluxetable}

%% file: tab3.tex
\begin{deluxetable}{ccccc}
\tablecaption{Temperatures and Column Densities \label{tab3}}
\tablewidth{0pt}
\tablehead {\multicolumn{3}{c}{H$_2$} & \multicolumn{2}{c}{FeII $/$ SiII}\\
\hline\\
\colhead{Position} & \colhead{T(K)}  &  \colhead{N(H$_2$) (cm$^{-2}$)} & \colhead{T(K)} & \colhead{n$_e$(cm$^{-3}$)}}
\startdata
CS & 900$\pm$170 & 3.18 10$^{17}$   & \nodata & \nodata  \\
CN & 755$\pm$35 & 7.81 10$^{17}$   & $<$1700 & 200-1000  \\
OF1 & 600$\pm$30 & 1.92 10$^{18}$   & \nodata & \nodata \\
OF2 & 670$\pm$50 & 2.34 10$^{18}$   & $<$2500 & 300-500  \\
\enddata
\end{deluxetable}

%% file: tab4.tex
\begin{deluxetable}{cccc}
\tablecaption{Gas-phase abundance of Fe and Si \label{tab4}}
\tablewidth{0pt}
\tablehead {\colhead{Position}  &  \colhead{CN} & \colhead{OF1} & \colhead{OF2}}
\startdata
${[Fe_{gas}]/[Fe_{\odot}]}^a$    & 3-5\,10$^{-2}$   & 4-7\,10$^{-2}$ & 3-5\,10$^{-2}$ \\
${[Fe_{gas}]/[Fe_{\odot}]}^b$    & 1-2\,10$^{-1}$   & 1.5-3\,10$^{-1}$ & 1-2\,10$^{-1}$ \\
${[Si_{gas}]/[Si_{\odot}]}^a$   &  3-5\,10$^{-2}$& \nodata &3-5\,10$^{-2}$  \\
${[Si_{gas}]/[Si_{\odot}]}^b$   & 1-2\,10$^{-1}$& \nodata &1-2\,10$^{-1}$  \\

\enddata
\\
~$^a$ assuming collisions with electrons and n$_e$ =400 cm$^{-3}$\\
~$^b$ assuming collisions with n$_H$, n$_{tot}$=10$^5$ cm$^{-3}$ and n$_H$ = 0.1 n$_{tot}$
\end{deluxetable}
%

%% file: tab5.tex
\begin{deluxetable}{cccc}
\tablecaption{Physical parameters \label{tab5}}
\tablewidth{0pt}
\tablehead {\colhead{Position}  &  \colhead{\.{M}(H$_2$)} & \colhead{\.{M}([SI])} & \colhead{\.{M}([FeII])}\\
  & \multicolumn{3}{c}{(10$^{-7}$M$_{\odot}$\,yr$^{-1}$)}}
\startdata
CN &  0.5  &9--20  & 2--20\\
OF2 &  1.4  &  3--8 & 0.9--9 \\
\enddata
\end{deluxetable}